\documentclass{article}
\usepackage[utf8]{inputenc}
\usepackage{graphicx}
\usepackage{booktabs}
\usepackage{authblk}
\usepackage{url}
\title{Conceptual Design of a Polarized $^3$He Target \\for the CLAS12 Spectrometer}
\author[1]{James Maxwell}
\author[2]{Richard Milner}
\affil[1]{Jefferson Lab, Newport News, VA}
\affil[2]{Laboratory for Nuclear Science, MIT, Cambridge, MA}

\begin{document}

\maketitle
\begin{abstract}
We present a conceptual design for a polarized $^3$He target for Jefferson Lab's CLAS12 spectrometer in its standard configuration. This two-cell target will take advantage of advancements in optical pumping techniques at high magnetic field to create 60\% longitudinally polarized $^3$He gas in a pumping cell inside the CLAS12 5\,T solenoid. By transferring this gas to a 20\,cm long, 5\,K target cell, a target thickness of $3 \times 10^{21}$ $^3$He/cm$^2$ will be produced, reaching the detector's specified maximum luminosity with a beam current of 2.5\,$\mu A$. 

\end{abstract}

\section{Introduction}
\begin{figure}[h!]
\centerline{\includegraphics[width=0.9\textwidth]{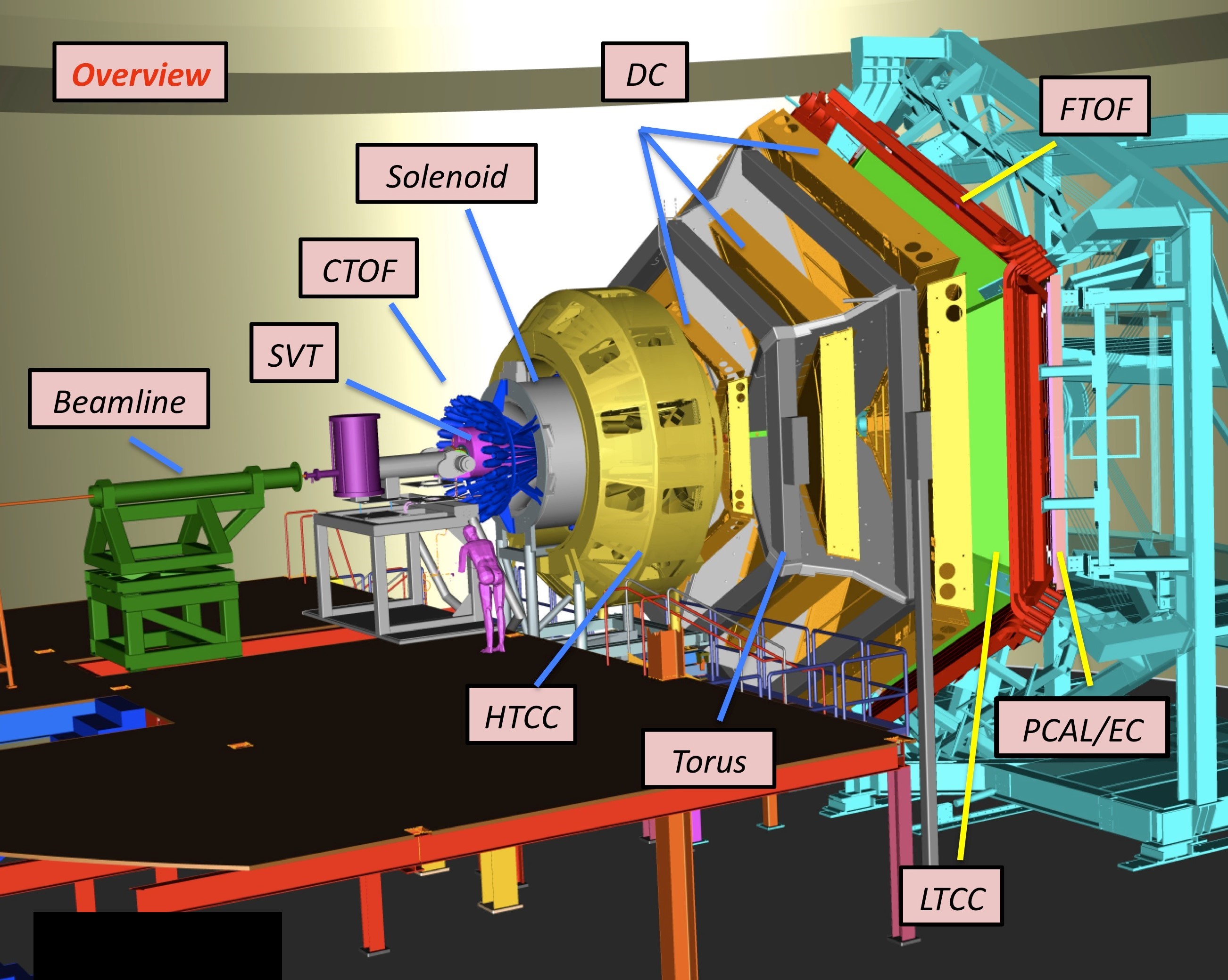}}
\caption{Schematic layout of the CLAS12 spectrometer.}
\label{fig:CLAS12}
\end{figure}
The CLAS12 spectrometer, shown in Figure~\ref{fig:CLAS12} in Hall B at Jefferson Lab is a powerful tool employed by the CLAS12 collaboration  to investigate nucleon and nuclear structure using the 11 GeV polarized electron beam delivered from the recently upgraded CEBAF.  It has been carefully designed~\cite{CLAS12TDR} to measure the complete electromagnetic response for nucleons and nuclei across the kinematic plane: elastic, resonance, quasielastic and deep inelastic reactions (inclusive, semi-inclusive and deeply virtual processes). It possesses many attractive characteristics:
\begin{itemize}
    \item large acceptance
    \item high luminosity, up to $10^{35}$ nucleons/cm$^2$/s
    \item excellent particle identification: $e/p/\pi/K/\gamma,...$
    \item neutron detection
    \item tagged particle detection.
\end{itemize}
Spin-dependent electron scattering from polarized $^3$He is well established as a means to provide an effective polarized neutron target.  Thus, a polarized $^3$He target in CLAS12 would make accessible a wide class of new observables that can provide additional insight into the quark and gluon structure of the neutron. Further, the few-body spin structure of polarized $^3$He offers the potential to measure the spin-structure of the proton and deuteron in the bound $^3$He nucleus~\cite{Milner2018}, if tagged events can be detected efficiently.  Motivated by these physics possibilities with a polarized $^3$He target located within the standard configuration of the CLAS12 spectrometer, we here develop a conceptual design for such a target to be located in the 5\,T central solenoidal magnet.

\subsection{Existing Polarized $^3$He Target Technology and CLAS12 Constraints}
Development of polarized $^3$He target technology has made great progress since the late 1980's.  Gas targets employing both Spin Exchange Optical Pumping (SEOP) and Metastability Exchange Optical Pumping (MEOP) have been employed for scattering experiments in nuclear and particle physics at MIT, TRIUMF, IUCF, SLAC, DESY, Mainz and Jefferson Lab.  See~\cite{Gentile2017} for a recent, general review.  For example at Jefferson Lab, a beam current of 30 $\mu$A on a 40 cm long target at 10 atm, yielding a luminosity of $2.2 \times 10^{36}$ $^3$He/cm$^2$/s with target polarization of 55\%, has been recently reported~\cite{Chen2019}.  However, all of these gaseous targets are polarized via optical pumping at {\bf low} magnetic fields.

The standard configuration of the CLAS12 spectrometer contains a central 5\,T solenoidal magnet which is used to track final-state particles. A conventional polarized $^3$He target can only be used with CLAS12 with one of the following major modifications:
\begin{itemize}
\item[1.] the central solenoid is removed
\item[2.] the polarized $^3$He target is located upstream of the CLAS12 spectrometer
\item[3.] the atoms are polarized in a low magnetic field and transferred into the 5\,T field of the central solenoid 
\end{itemize}
(1) and (2) have the consequence of requiring a major modification in the configuration of the CLAS12 spectrometer, which we wish to avoid. We have studied (3) extensively~\cite{Maxwell2015} and it is challenging to accomplish this without significant loss of polarization.

\section{Proposed Concept for a New Target}
Rather, we propose for the first time a gaseous polarized $^3$He target based on {\bf high}-field MEOP, a technique invented at the Laboratoire Kastler Brossel, {\'E}cole Normale Sup{\'e}rieure (ENS), Paris, France~\cite{Gentile2017}. In our proposed scheme, the atoms are directly polarized in the high field of the central solenoid and transferred diffusively to a cold target cell, through which the electron beam passes.  We have utilized this high-field MEOP in the development of a new polarized $^3$He ion source for RHIC~\cite{Max2019} and describe it here.

\subsection{High-Field MEOP}
\begin{figure}[!p]
\centerline{\includegraphics[width=0.9\textwidth]{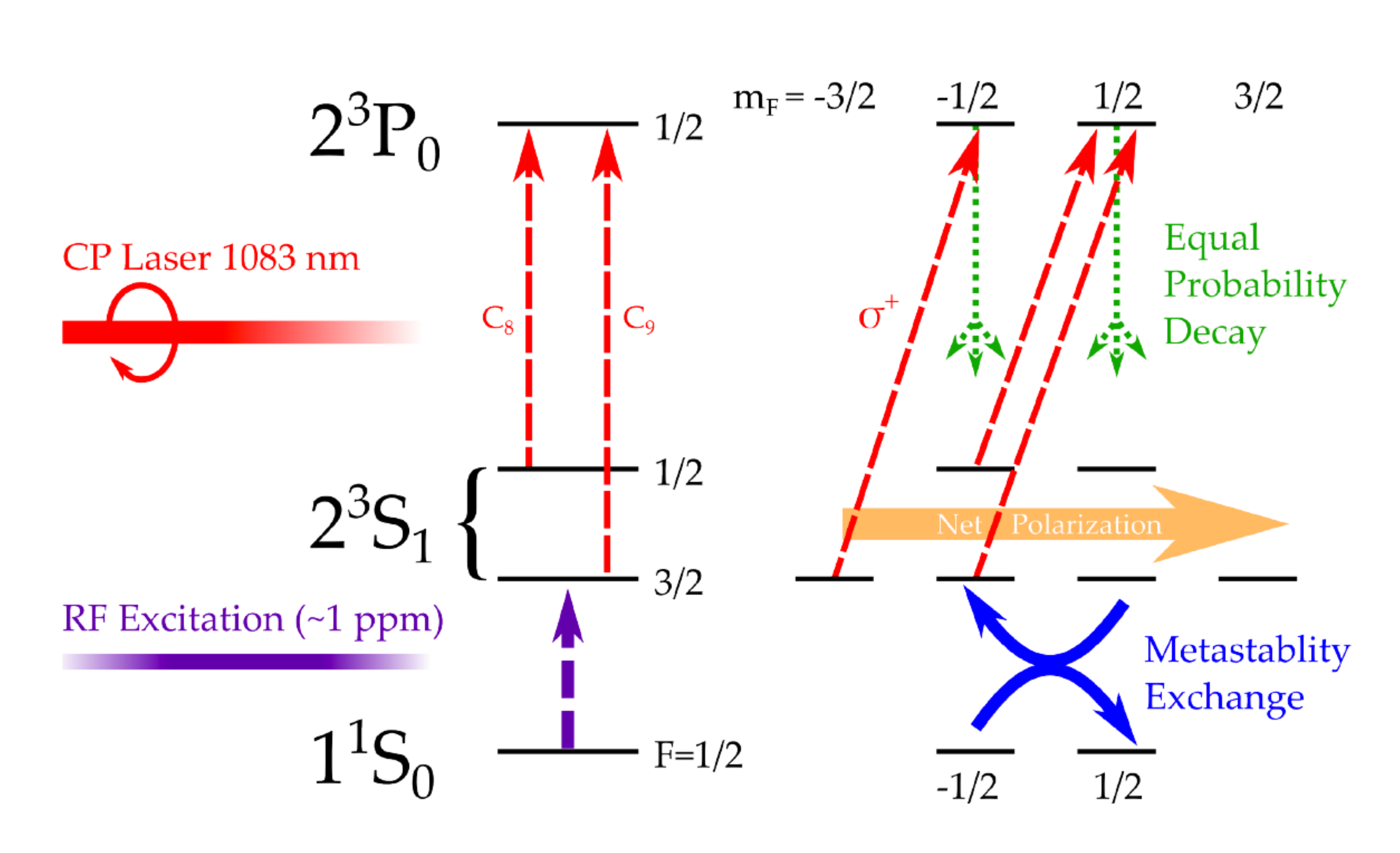}}
\caption{Atomic states of $^3$He and the metastability exchange mechanism.}
\label{fig:MEOP}
\end{figure}

The MEOP technique was invented by Colegrove, Schearer and Walters in 1963~\cite{Cole1963}. Figure~\ref{fig:MEOP} shows the atomic states of the $^3$He atom.  An RF discharge excites $\sim$1 ppm of $^3$He gas into the 2$^3$S metastable state.  This state is optically pumped using circularly polarized laser light at 1083 nm to the excited 2$^3$P state and the metastable state becomes polarized.  Subsequently, the metastability exchange process transfers polarization to the nucleus of the ground state atom.  The presence of the discharge also gives rise to depolarization processes that limit the maximum polarization achievable. 

\begin{figure}[!p]
\centerline{\includegraphics[width=0.9\textwidth]{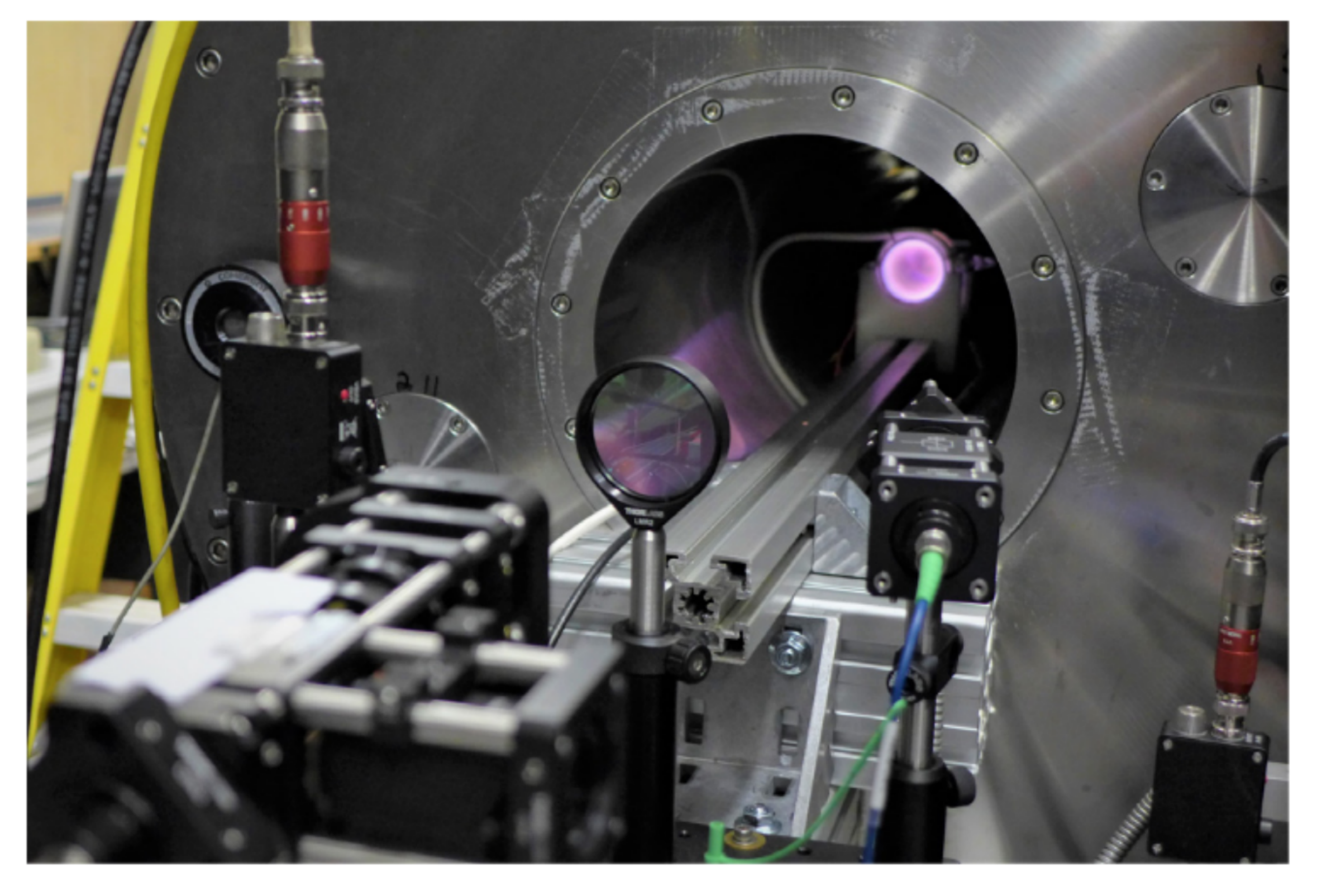}}
\caption{Photograph from~\cite{Max2019} of the polarizing apparatus and EBIS spare solenoid warm bore at BNL. In the foreground are the pumping laser circular polarization optics. The probe laser fiber enters a circular polarizer on the right, and after passing through the cell the probe light is reflected by a mirror back to a photodiode on the left. The sealed cell is illuminated by the RF discharge plasma in pink; for this photograph it is much brighter than is effective for optical pumping.}
\label{fig:Photo}
\end{figure}

In high magnetic fields, Zeeman splitting reduces coupling between electron and nuclear spins, making MEOP less efficient. However, this splitting also inhibits polarization relaxation channels and separates hyperfine states so that they may be cleanly addressed by the high-power lasers now available, ultimately permitting high steady-state polarization.  Further, high-field MEOP becomes increasingly more efficient at higher gas pressures, at which low-field MEOP is not tenable. Figure~\ref{fig:Photo} is a photograph of the high-field apparatus as used at the EBIS solenoidal magnet at BNL.

\begin{figure}[!h]
\centerline{\includegraphics[width=0.8\textwidth]{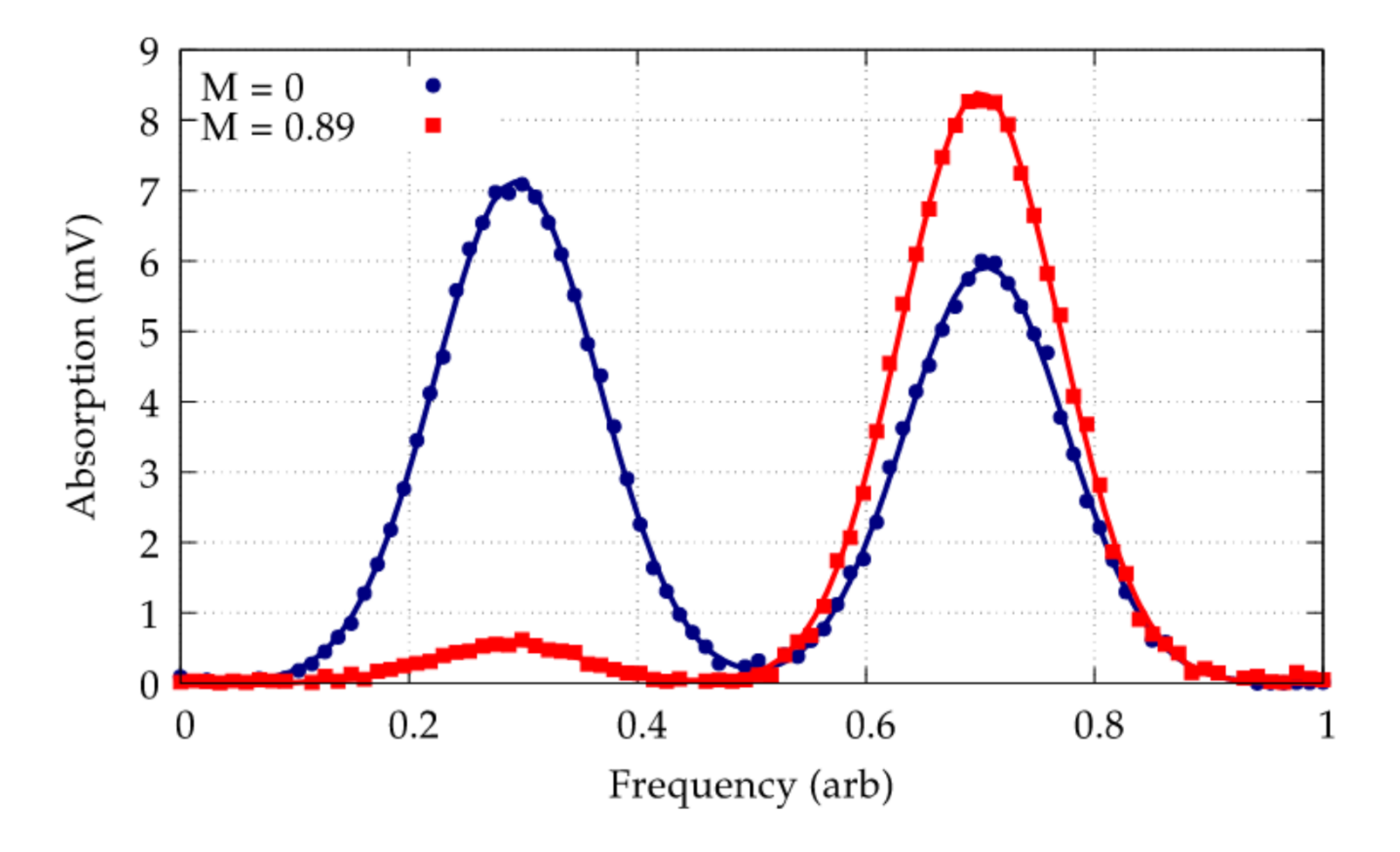}}
\caption{Absorption signals for populations $a_1$ (left) and $a_2$ (right) measured  with ground state nuclear polarization at 0 and 89\%, using a 1 torr sealed cell at 3 T~\cite{Max2019}.}
\label{fig:NIMpol}
\end{figure}

At high magnetic field, the population of two particular 2$^3$S sub-levels can be monitored by sweeping the probe laser frequency, providing a measure of the ground state polarization~\cite{Gentile2017}. These sublevels are chosen to avoid the states under active pumping for polarization.  At spin-temperature equilibrium, the populations of these probed states, here $a_1$ and $a_2$ will satisfy $a_2/a_1 = e^{\beta} = (1+M)/(1-M)$.  An absolute measure of the nuclear polarization $M$ of the ground states can be formed from the change in the ratio $r=a_2/a_1$ of the absorption signal amplitudes for these sublevels during MEOP, as calibrated by their ratio $r_0$ when not polarized ($M=$ 0):
\begin{equation}
M = \frac{r/r_0-1}{r/r_0+1} \ .
\end{equation}
Because only ratios of spectral amplitudes are involved, all experimental parameters affecting the absolute signal intensities are canceled out, making this a robust measurement.  To measure the polarization in our sealed $^3$He cells, we built an optical probe polarimeter in the style of the ENS group~\cite{Nikiel2013}.  Figure~\ref{fig:NIMpol} shows sample absorption signals for a high value of polarization.

\begin{figure}[!hp]
\centerline{\includegraphics[width=0.8\textwidth]{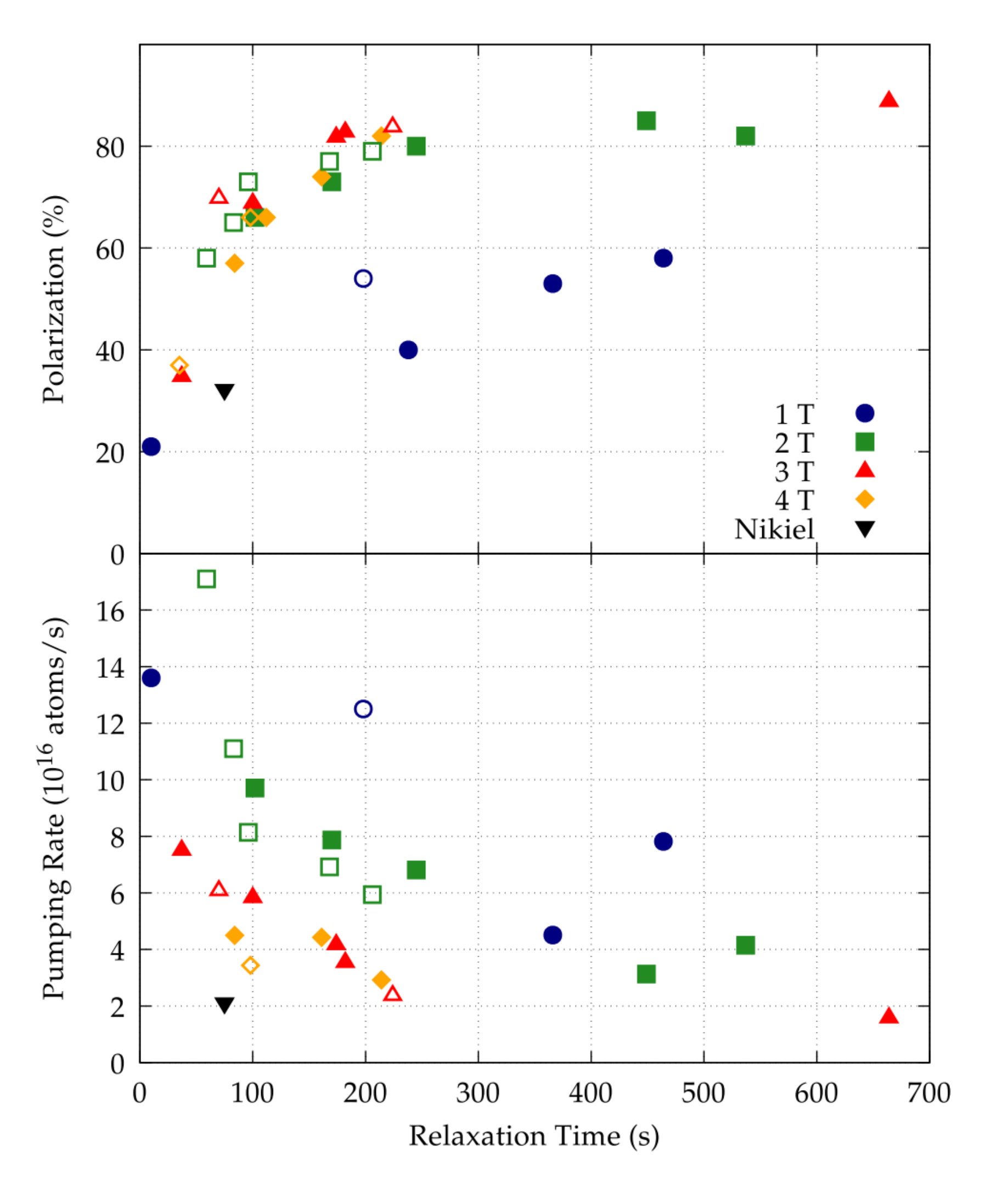}}
\caption{Steady state polarization achieved (top) and corresponding pumping rate (bottom) for a given relaxation time with the plasma discharge. Different shapes represent four magnetic field settings from 1 to 4 T. Filled shapes designate measurements on a 1 torr sealed cell produced a MIT Bates, while open shapes designate those taken on a 1 torr sealed cell on loan from T. Gentile of NIST. A single 1 torr, 4.7 T result from Nikiel \textit{et al}~\cite{Nikiel2013} is also shown.}
\label{fig:NIMdata}
\end{figure}

The upper panel of Figure~\ref{fig:NIMdata} shows our achieved, steady-state nuclear polarizations at various $T_D$ relaxation times. Here the relaxation time is largely a function of discharge intensity, with the dimmest discharges resulting in the longest relaxation times. All these measurements were
performed on two 1 torr sealed cells.  As has been noted by others, we do not see a strong steady-state polarization dependence on laser power in the 1 to 4 W range, and more power than this tends to be counter-productive.
With the exception of the 1 T setting, we do not see a strong dependence on the magnetic field. At 2, 3 and 4 T, polarizations exceeding 80\% were seen, even with relatively short relaxation times below around 300 s. 

\begin{figure}[h]
\centerline{\includegraphics[width = 0.8\textwidth]{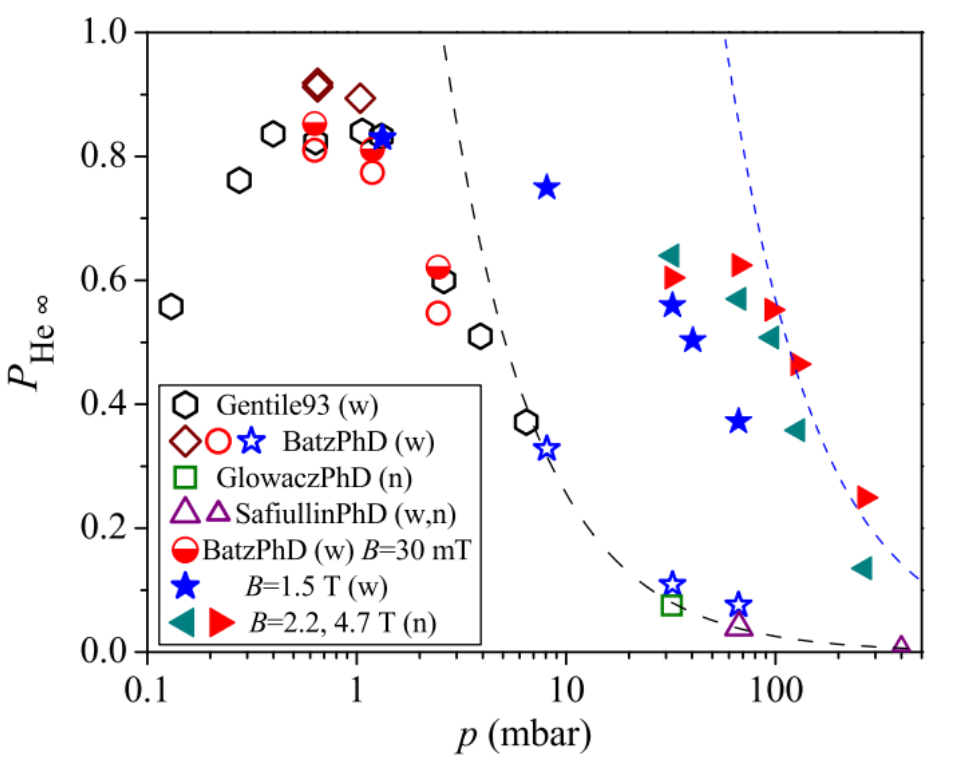}}
\caption{Variation with pressure of highest steady-state polarizations achieved by various groups at low fields (open symbols, 1 to 3 mT) and high fields (filled symbols, see the legend) from~\cite{Gentile2017}.}
\label{fig:ENS}
\end{figure}

The ENS group has undertaken extensive study of the improvement of MEOP efficiency at pressures from 1 to 300\,mbar with increasing magnetic field. Figure \ref{fig:ENS} shows the achieved steady-state polarization versus gas pressure for magnetic fields up to 4.7\,T. With no further improvement in the technique, polarizations as high as 60\% are possible at 100\,mbar, two orders of magnitude higher than typical MEOP pressures.

\subsection{Double-Cell $^3$He Target System}
\begin{figure}[h!]
\centerline{\includegraphics[width=0.9\textwidth]{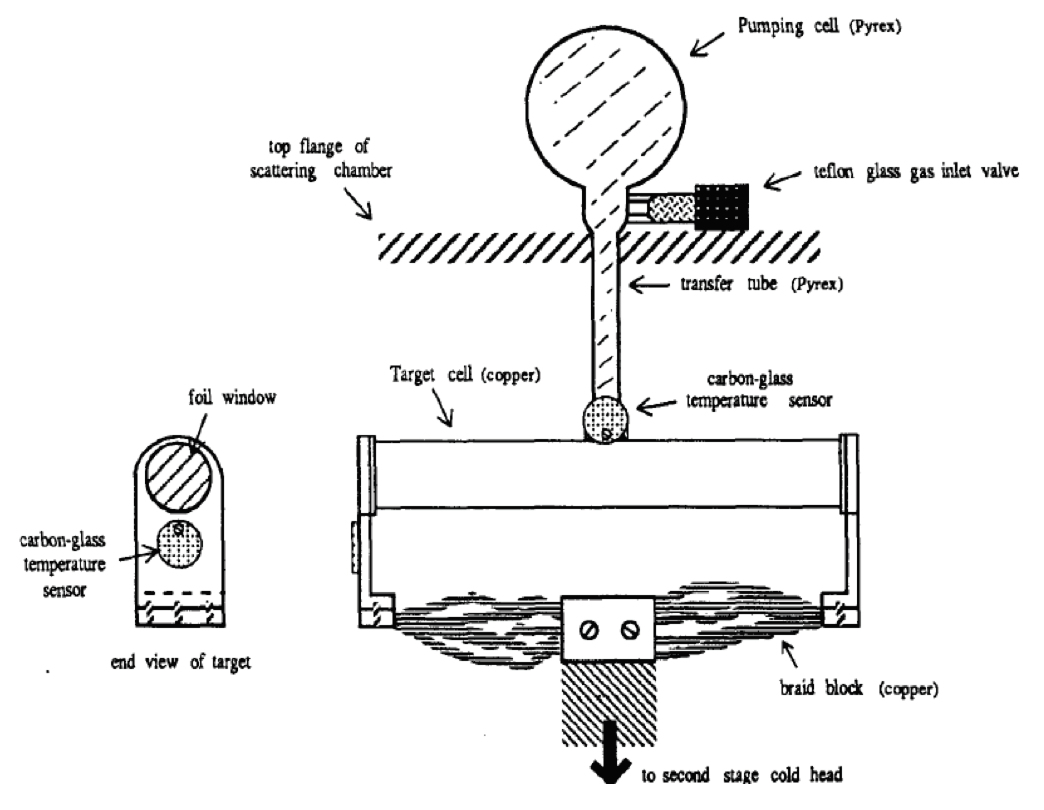}}
\caption{Schematic layout of the polarized $^3$He target double-cell system developed at Caltech~\cite{Milner1989,Jones1992} for Bates experiment 88-02.}
\label{fig:CIT-targ}
\end{figure}

A double-cell polarized $^3$He target was developed~\cite{Milner1989,Jones1992} at Caltech for Bates experiment 88-02~\cite{Batesprop1987} and was the first such target used for electron scattering experiments, including spin-dependent inclusive scattering in the quasielastic region~\cite{Wood1990}. It used an LNA-based, custom-built laser system which has been superseded by modern, commercially available, turn-key fiber-based systems.  The copper target cell had a circular cross section of diameter 2.54 cm and a length of 16 cm.   The interior of the target cell was coated with a thin layer of frozen nitrogen to reduce depolarization from interactions with the cell walls. The end windows were 4.6 $\mu$m thick copper foils which were epoxied to the target cell.  Fig.~\ref{fig:CIT-targ} shows a schematic layout of the Bates 88-02 target. 

During data taking, a total integrated charge of 1478 $\mu$A-hours was accumulated on this target. It was subsequently used in a second set of measurements~\cite{Gao1994,Hansen1995} of inclusive quasielastic spin-dependent electron scattering from polarized $^3$He at Bates in 1993.

\subsection{Measuring Polarization in a Double-Cell Target}

In the system proposed here, we intend to use the polarization measurement technique developed for the Bates 88-02 target, namely inferring the polarization of the gas inside the target cell based on the measured polarization in the pumping cell, as described in~\cite{Milner1989,Jones1992}. By solving simple rate equations which describe the rate of polarization relaxation in each cell and the rate of diffusion between the cells, the target cell polarization can be determined.

The expressions used to extract the target cell polarization from the measured polarization in the pumping cell depend strongly upon the time constants of the system. The most relevant of these are the relaxation times in the pumping cell and the target cell, and the time for atoms to transfer between the two volumes. 
The spin relaxation rate of the system is measured by shuttering the laser light so that the atoms are no longer being optically pumped and observing the decay of the polarization as a function of time. For a single cell, the polarization decays exponentially, 
\begin{equation}
P(t) = P(0)e^{-t/\tau}
\end{equation}
with a single time constant, $\tau$, parameterizing the rate. The time constant reflects the combination of all the effects contributing to the spin relaxation. For example, if there are two depolarization mechanisms contributing to the relaxation, which are individually characterized by the time constants $\tau_1$ and $\tau_2$, then the overall relaxation rate is 
\begin{equation}
\frac{1}{\tau} = \frac{1}{\tau_1} + \frac{1}{\tau_2} \ .
\end{equation}
The single exponential form of the expression for the spin relaxation process is altered for a system containing more than one cell.
\begin{figure}[h!]
\centerline{\includegraphics[width=0.8\textwidth]{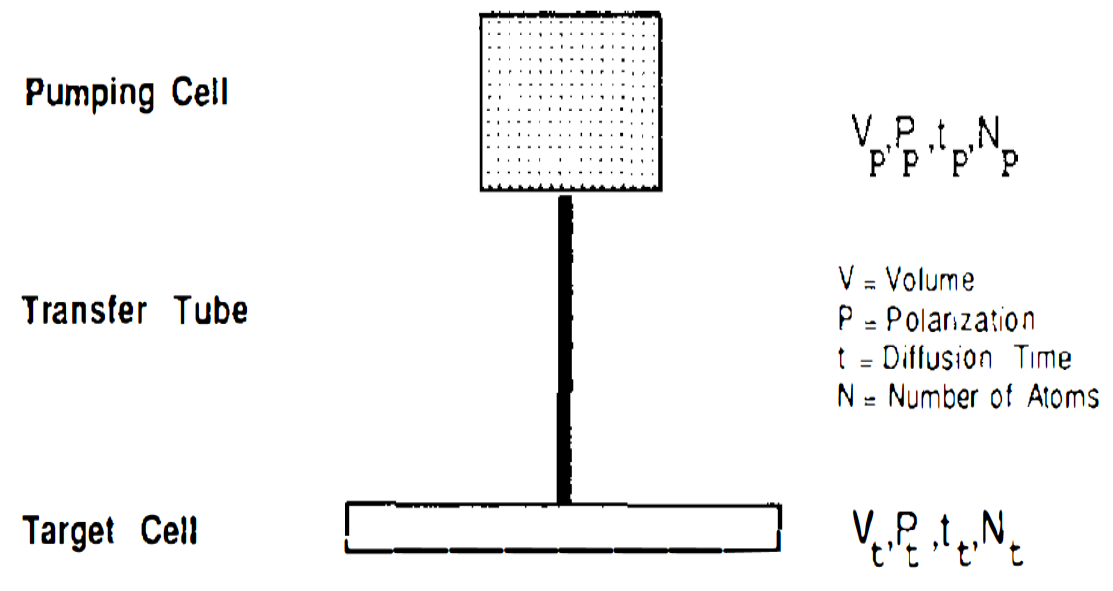}}
\caption{Schematic diagram of the double cell target.}
\label{fig:Twocell}
\end{figure}

Fig.~\ref{fig:Twocell} shows a schematic diagram of a double cell system. We denote by $V_p$, $P_p$, $N_p$ and $t_P$ the volume, polarization, number of atoms and average sitting time of an atom in the pumping cell.  Similarly, we have quantities $V_t$, $P_t$, $N_t$, and $t_t$ for the target cell. The total number of atoms in the double cell system is $N = N_P + N_t$. The volume of the capillary tube is neglected as it amounted to only 2\% of the volume of the target cell. From the known volumes of the pumping and target cells and the measured target temperature, $N_P$ and $N_t$ can be determined. Also, in a steady state condition we have
\begin{equation}
\frac{N_p}{t_p} = \frac{N_t}{t_t} \ .
\end{equation}

The polarization of the target cell may be determined in terms of the polarization of the pumping cell by observing the pumping cell polarization as it relaxes without the influence of optical pumping. One technique uses adiabatic fast passage NMR (AFP) to reverse the polarization of the atoms in the pumping cell, $P-p$, without affecting the polarization of the atoms in the target cell, $P_t$. The double cell system is first polarized with the laser, then the optical pumping light is shuttered and the pumping cell spins reversed by AFP. The reversed spins in the pumping cell come into diffusive equilibrium with the unchanged spins in the target cell. This occurs with a time constant given by the communication time between the two cells 
\begin{equation}
\frac{1}{t} = \frac{1}{t_p} + \frac{1}{t_t} .
\end{equation}
 The time constants are defined in the relaxation rate equations for the coupled two-cell system. 
\begin{eqnarray}
\frac{dP_p}{dt} &=& \frac{-P_p}{\tau_p} + \frac{P_t-P_p}{t_p} \\
\frac{dP_t}{dt} &=& \frac{-P_t}{\tau_t} + \frac{P_p-P_t}{t_t} 
\end{eqnarray}
The subscript ``t"(``p") refers to the target(pumping) cell. $\tau_{t(p)}$ is the relaxation time in the cell and $t_{t(p)}$ is the diffusion time for the population of atoms in the cell to transfer out. 

For a double-cell system, the decay of the polarization is expressed as a sum of two exponentials. In a system designed so that the transfer rate between the two cells is much faster than the relaxation rate in either cell, the decay is made up of one exponential with a short time constant, which is approximately the transfer time between the two cells, $t_{ex}$ , and a second exponential with a long time constant, which is approximately the weighted average of the relaxation time constants for the two cells, where the weighting factor is the fraction of the atoms in each cell. 

\begin{figure}[h!]
\centerline{\includegraphics[width=0.8\textwidth]{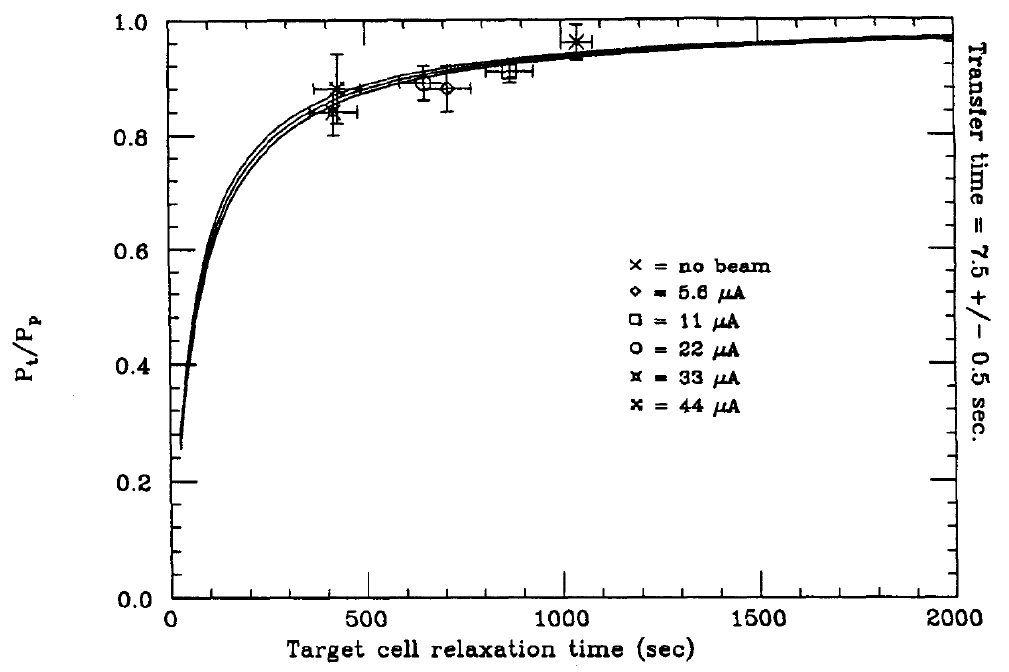}}
\caption{$P_t/P_p$ as a function of the relaxation time in the target cell, for transfer times of $t_{ex}$ = 7.5 $\pm$ 0.5 sec, shown with the data from the study of beam depolarization with a minimum ionizing electron beam~\cite{Jones1992}.  The beam currents at which the data were taken are indicated.}
\label{fig:Targpol}
\end{figure}

Information from the polarizing curves is used to calculate the ratio of the polarizations in the two cells under equilibrium conditions, i.e., after the optical pumping light has been on the pumping cell for long enough that steady state conditions have been established in the cells. The premise underlying this indirect measurement to determine the polarization of the target cell nuclei is that the pumping cell polarization relates to the target cell polarization in a straightforward manner. The equilibrium ratio of the polarizations of the two cells, $P_t/P_p$, is found to depend only upon the relaxation time in the target cell, the transfer time, and the fraction of atoms in each cell 
\begin{equation}
\frac{P_t}{P_p} = \left[1+\frac{N}{N_p} \frac{t_{ex}}{\tau_t} \right]^{-1} \ .
\end{equation}
Once these constants have been measured, the target cell polarization is simply calculable from the pumping cell polarization. One thing that should be noted is that the equilibrium condition between the two cells must be met before the simple ratio holds. 
Fig.~\ref{fig:Targpol} shows the calculated equilibrium polarization ratio for the Caltech target.
For the target system developed at Caltech, equilibrium was established in about 8-10 minutes for pumping the system from an initially unpolarized state and much more quickly for changes in the relaxation time brought about by changes in the beam current.

The extraction of the target cell polarization depends upon the transfer time and the relaxation time in the target cell. The transfer time is well determined because it does not vary significantly between measurements, depending only upon the temperature and pressure, which are held almost constant throughout the experiment. However, the target cell relaxation time depends upon the surface conditions and the amount of beam current, which change with time and need to be measured. The uncertainty in the relaxation time is a major source of uncertainty in the extraction of the target polarization. Also, there may be gradients in the $^3$He polarization in the target cell due to beam effects.
The target cell relaxation time was typically about 1000 sec.

\subsection{Proposed CLAS12 Target}
We propose that a two-cell target system, similar to the target developed at Caltech for Bates experiment 88-02~\cite{Milner1989,Jones1992}, be installed in the 5T magnetic field of the CLAS12 central solenoid and using a pumping cell pressure of 100 mbar. This system would share many design elements of the liquid target currently under development for CLAS12, utilizing liquid helium to cool the $^3$He target cell to 5\,K.
The available space for the complete target system in the CLAS12 central solenoid is estimated to be a cylinder of diameter 10 cm and length 47 cm with the symmetry axis being the direction of the incident electron beam. 

\begin{figure}[t]
\centerline{\includegraphics[width=1.1\textwidth]{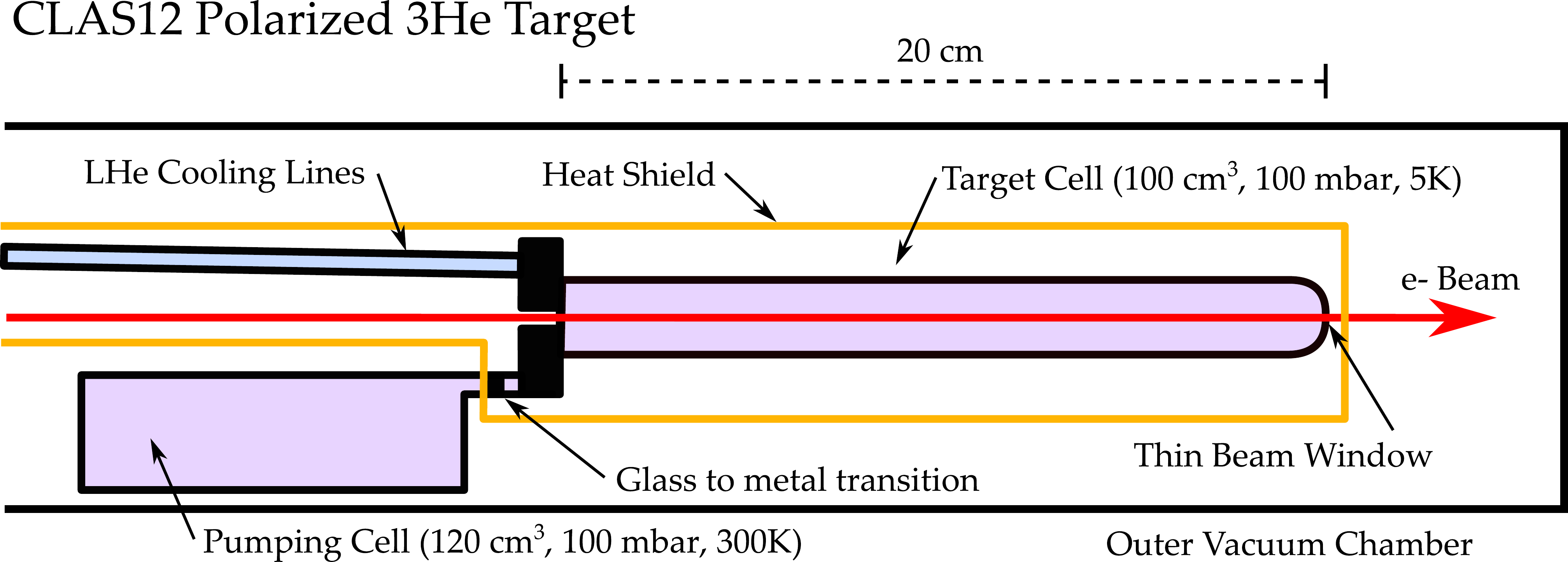}}
\caption{Schematic layout of a two cell polarized $^3$He target system in CLAS12.}
\label{fig:Sideview}
\end{figure}

The glass pumping cell would be at room temperature and would be in diffusive contact with a target cell constructed of aluminium maintained at 5 K to increase the gas density by a factor of $300/5 = 60$.  The ENS group has achieved 60\% polarization at a pressure of 100 mbar. At 100\,mbar, we would reach a gas density of 5.4\,amg at 5\,K, which compares favorably to the 9\,amg typically used at low field and room temperature by Spin-Exchange Optical Pumping targets in Hall A.
Fig.~\ref{fig:Sideview} shows a schematic layout that fits in the available space.  The glass pumping cell and the target cell are both cylindrical with diameters of 4 and 2.5\,cm, and lengths of 10 and 20 cm, respectively.

Table~\ref{tab:Comp} compares the achieved specification for the Bates 88-02 target with those proposed for the CLAS12 target based on demonstrated high-field optical pumping performance. We note that our proposed target yields a luminosity per nucleon surpassing the maximum value of $10^{35}/\textrm{cm}^2/\textrm{s}$ specified by the CLAS12 design.

\begin{table}[h]
\begin{center}
\begin{tabular}{lcc}
\toprule
 &  Bates 88-02   &  CLAS12 \\
 Parameter         &   Target   &  Target  \\
          & Achieved & Proposed\\
          \midrule
Pumping cell pressure (mbar) & 2.6 & 100\\
Pumping cell volume (cm$^3$) & 200& 120\\
Target cell volume (cm$^3$) & 79& 100\\
Target cell length (cm) & 16& 20\\
Number of atoms in pumping cell &$ 1.2 \times 10^{19}$ & $3 \times 10^{20}$\\
Number of atoms in target cell & $6 \times 10^{19}$ & $1.5 \times 10^{22}$\\
Holding field (T)  &0.003 & 5\\
Polarization & 40\% & 60\% \\
Incident electron beam energy (GeV) & 0.574 & 10\\
Cell temperature (K) & 17 & 5\\
Target thickness ($^3$He/cm$^2$) &$1.2 \times 10^{19}$ & $3 \times 10^{21}$\\
Beam current ($\mu$A) & 10 & 2.5\\
Luminosity ($^3$He/cm$^2$/s) & $7.2 \times 10^{32}$& $4.5 \times 10^{34}$\\
\bottomrule
\end{tabular}
\end{center}
\caption{Comparison of specifications for the Bates 88-02 target~\cite{Jones1992} and the CLAS12 polarized $^3$He target.}
\label{tab:Comp}
\end{table}

\section{Path to Realization}

\begin{figure}[h!]
\centerline{\includegraphics[width=0.9\textwidth]{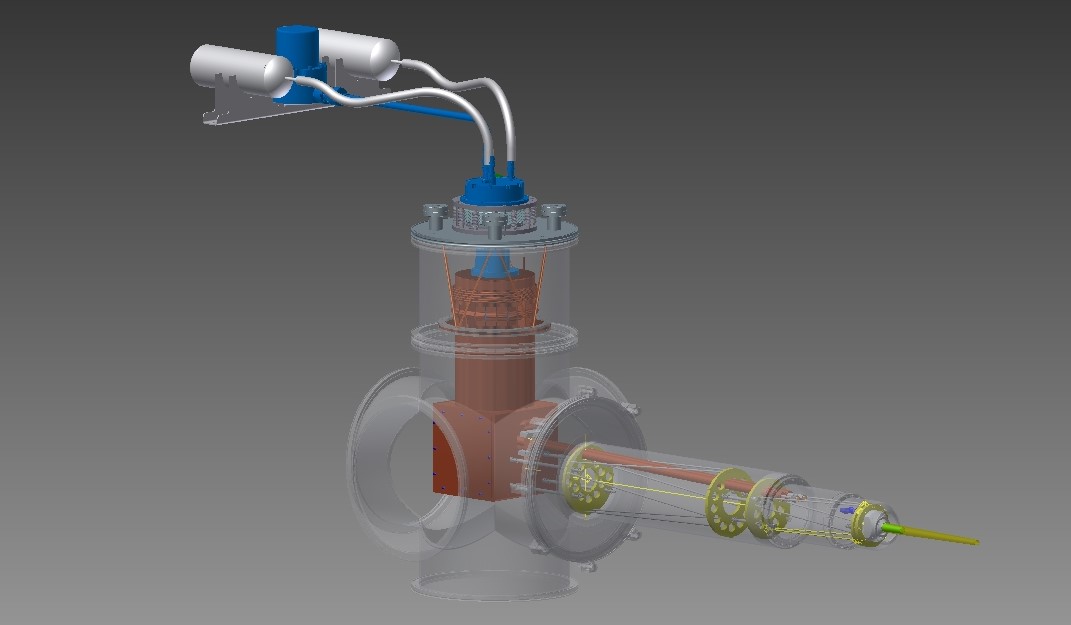}}
\caption{Rendering of the Hall D cryotarget, which condenses liquid hydrogen or helium using a pulse tube cryocooler. }
\label{fig:HallD}
\end{figure}

A new cryotarget is currently under development to provide liquid targets for CLAS12, in the style of the Hall D cryotarget shown in Figure \ref{fig:HallD}. This system will use a Cryomech PT-420 pulse tube, which can provide 2\,W of cooling power at 4.2\,K, to condense hydrogen or helium to supply a target cell within the CLAS12 solenoid. By instead recirculating the condensed liquid helium through a heat exchanger on a gas target cell, this design could provide sufficient cooling power for a 5\,K polarized $^3$He gas target cell.  We are pursuing a design which would utilize identical cryogenic infrastructure as---or perhaps even act as a modular addition to---the new Hall B cryotarget, where a polarized gas cell would take the place of the liquid cell.

Care must be taken to design the system within the tight space constraints of the CLAS12 solenoid and silicon vertex tracker. The $^3$He gas target will require a separate pumping cell, gas supply lines, pump and probe laser fibers, and a heat shield that will surround the target cell but exclude the pumping cell. A number of other issues will require special consideration in the development of a full-scale prototype.


\subsection{Target Cell Design}
Wall depolarization was systematically studied~\cite{Jones1992} by the Caltech group in the design of the two-cell target.  Coating the cold target cell wall with frozen nitrogen worked well in the Bates 88-02 target to deliver long relaxation times ($>$ 500 sec). Without a wall coating, the measured target cell wall relaxation time was 1 $-$ 2 orders of magnitude shorter at 13.5 K. Other groups have reported that cryogenic H$_2$ coatings have yielded days-long relaxation times when used in the range of 2 to 6\,K \cite{Gentile2017,Lefevre1988}. 

The cell will be designed to avoid occluding CLAS12's acceptance. We propose cells similar to those used the the Hall A tritium experiments, with the massive ``lid'' on the upstream end containing the heat exchanger, and a thin-walled aluminum shell forming the main volume of the cell. The cell walls need only be thick enough to safely contain the gas and allow conduction of beam heat from the downstream window.

The end windows used for the 2\,mbar Bates 88-02 target were 4.6 $\mu$m thick copper foils epoxied to the target cell. While it is desirable to make the end windows as thin as possible to reduce background, the increased pressure in will likely require aluminum beam windows of at least 200 $\mu$m.

\subsection{Heat Load}

The power deposition due to energy loss by the beam in the target gas will be roughly 100\,mW, while the foil windows are estimated contribute an additional 7 mW. The glass transfer tube between the cells will also conduct heat to the target cell, creating another roughly 30\,mW of heat to remove. 

As MEOP does not operate effectively at low temperature, the pumping cell will be kept at room temperature. A simple calculation of heat exchange via conduction in a static gas gives roughly 35\,mW of heat. To look for an upper limit on this heat load, we can assume that all the gas in the pumping cell is cooled to 5\,K every 8 seconds (an estimated transfer time from the Bates target and overestimate at these pressures), which would result in roughly 350 mW of heat. 

The transfer line between the cells will need to be optimized to balance the heat load from gas diffusion and the supply of polarized gas to the target cell, perhaps using the temperature differential between cells to create convective flow through two transfer lines. All together, these heat loads are within the 2\,W of cooling power of the system at 4.2\,K, however a detailed heat analysis will be performed before the design is finalized.

\subsection{Beam Depolarization}

Beam depolarization in the two-cell system was studied extensively by the Caltech group using beam at the Caltech Pelletron~\cite{Milner1989} and at MIT-Bates~\cite{Jones1992}.  From~\cite{Jones1992}, the contribution from beam depolarization to the target cell relaxation time was determined to be about 2000 sec at a beam current of 5 $\mu$A.  While the proposed beam current of 2.5 $\mu$A is lower, the proposed CLAS12 target density is up to 100 times higher than the Bates 88-02 target.  The diatomic molecular ion $^3{\rm He}^+_2$ provides the main mechanism for depolarization, and the ionization rate increases linearly with density~\cite{Singh2008}. 

A more detailed consideration is warranted, however, several approaches are available to ameliorate the effect. A key advantage of MEOP is the high rate of polarization, so beam depolarization could be counteracted simply by increasing the rate of communication between cells via convection. In addition, many of the reaction rates are not known at high magnetic field.  Further, the addition of a very small amount ($\sim10^{-4}$) of additional gas, like neon or nitrogen, acts to breakup the depolarizing diatomic molecule~\cite{Bonin1988}.  The ability to test a full prototype target system with low energy electron beam would be particularly valuable in understanding and minimizing the effects of beam depolarization. 


\subsection{Solenoidal Field Gradients}
The relaxation time $T_g$ due to transverse gradients in a holding field directed along the $z$-direction is given by
$$
\frac{1}{T_g} = \frac{<v^2>}{3}\frac{|\nabla B_x|^2 +|\nabla B_y|^2}{B^2_0} \left(\frac{\tau_c}{1+\omega_0^2 \tau_c^2} \right)
$$
where $B_0$ is the holding field, $\tau_c$ is the meantime between atomic collisions and $\omega_0 = \gamma B_0$ is the Larmor frequency for the magnetic field.  For $^3$He the gyromagnetic ratio, $\gamma$, is 3.24 kHz/G.  The mean collision rate has been measured as a function of pressure at 300 K and determined that
$$
\tau_c = (2.2 \pm 0.2) \times 10^{-7} p^{-1} \ {\rm sec} \ ,
$$
where $p$ is the pressure in Torr.  $<v^2> = 3kT/m$ is the mean square thermal velocity of the atoms.
The relaxation rate from magnetic field gradients decreases as the temperature decreases since the atoms move more slowly and therefore experience smaller fluctuations in the field for a given amount of time. In a double-cell system, where the target cell is cooled and the pumping cell is operated at room temperature, the effect of the field gradients is more important for the pumping cell.

\begin{figure}[h!]
\centerline{\includegraphics[width=\textwidth]{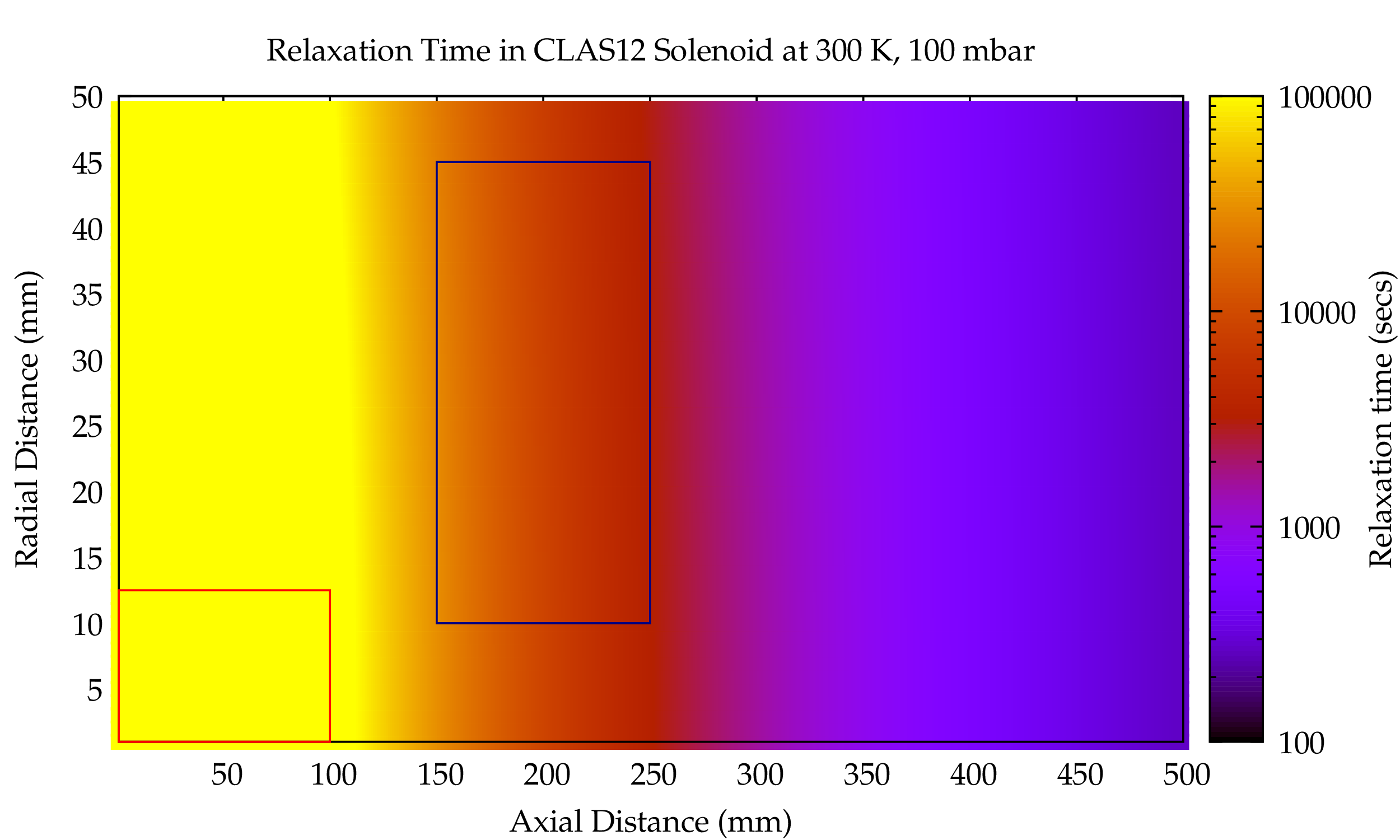}}
\caption{Preliminary map of $^3$He relaxation time due to transverse field gradients in the CLAS12 solenoid, showing distance from the center of the solenoid. This map assumes gas at 100\,mbar and 300\,K, and shows candidate locations for a pumping cell (blue box) and target cell (red box).}
\label{fig:map}
\end{figure}

For the pumping cell of the proposed CLAS12 target, we have
\begin{eqnarray}
B_0 &=& 5 \ {\rm Tesla} \\
{\rm pressure} &=& 75 \ {\rm Torr} \\
\tau_c &=& 2.2 \times 10^{-9} \ {\rm sec} \\
\omega_0 &=& 1.6 \times 10^8 \ {\rm Hz}\\
\end{eqnarray}
For a $T_g$ = 500 sec relaxation rate from transverse gradients, this yields
$$
\frac{|\nabla B_T|}{B_0} = 2 \times 10^{-3}\ {\rm per\ cm} \ .
$$
Thus, transverse gradients need to be less than 0.2\% per cm, which true throughout the central space of the solenoid.  Figure \ref{fig:map} shows a map of relaxation time in the central axial and radial space of CLAS12 for 300\,K and 100\,mbar $^3$He gas, showing candidate locations for pumping and target cells. For the 5\,K target cell, the gas particle velocity will be much lower, making relaxation times much higher than shown in this map. We expect that depolarization due to field gradients will be negligible if the pumping cell is located inside the solenoid.

\subsection{Beam Control and Dumping in Hall B}

The beam intensity of 2.5 $\mu$A required to reach maximum luminosity is significantly higher than current, typical operating currents in Hall B. At 10 GeV, the beam power is 25 kW. Modification of the beam dump and updates to radiation controls will likely be required.

\subsection{Transversely Polarized Target}
An important strength of JLab's SEOP $^3$He and solid DNP $p$ and $d$ targets has been the ability to polarize tranverse to the incident direction of the beam. Locating a transverse target inside the 5\,T longitudinal field of CLAS12 is problematic, but perhaps workable. The HD-Ice group has been actively investigating the use of bulk superconductor to both shield the longitudinal field and provide a transverse holding field~\cite{Statera2018}.
We have begun investigating the possibility adapting this scheme to a polarized $^3$He target.

\subsection{Prototype Construction}

To proceed, we propose that a full scale prototype target system, with dimensions consistent with the requirements imposed by the CLAS12 solenoid, should be designed, constructed and operated in a solenoidal magnetic field. JLab's new Upgraded Injector Test Facility could offer a convenient place to test a prototype system in a 5\,T warm-bore magnet to verify the target is operating as proposed under experimental conditions.

In addition, high-field optical pumping at pressures above 100 mbar should be investigated to see if a higher figure-of-merit can be attained. The ENS group is actively studying the limits to the polarization at high pressures. 
\section*{Summary}
The combination of novel, high-field MEOP techniques and a cryogenic target cell offers a new approach to high-luminosity scattering experiments on polarized $^3$He nuclei for applications requiring a high magnetic field, such as CLAS12. This scheme would provide a 20\,cm long, 60\% longitudinally polarized $^3$He target for use with a 2.5\,$\mu$A electron beam current, which would exceed the CLAS12 design luminosity of $10^{35}$ interactions per cm$^2$ per second.

\section*{Acknowledgements}

We thank Pierre-Jean Nacher and Genevieve Tastevin from the Laboratoire Kastler Brossel, {\'E}cole Normale Sup{\'e}rieure, Paris, France for their hospitality during the record heat in Paris in July 2019 and for sharing their data and insights into high field MEOP at high pressures.  We thank Cyril Wiggins and Marco Battaglieri from Hall B, Jefferson Laboratory, and Victoria Lagerquist from ODU for providing us with important technical information. We gratefully acknowledge Chris Keith, Dave Meekins and James Brock from the Jefferson Laboratory Target Group for several helpful discussions. This material is based upon work supported by the U.S. Department of Energy, Office of Science, Office of Nuclear Physics under contract DE-AC05-06OR23177.

\end{document}